\documentclass[usenatbib]{mnras}

\usepackage{graphicx}	
\usepackage{amsmath}	
\usepackage{amssymb}	
\usepackage{multicol}        
\usepackage{bm}		
\usepackage{pdflscape}	

\usepackage[T1]{fontenc}
\usepackage{ae,aecompl}

\usepackage{newtxtext,newtxmath}

\title[500 micron Risers I]{The Nature of 500 micron Risers I: SMA Observations}

\author[J.S. Greenslade et al.]{
J. Greenslade$^{1}$,
D.L. Clements$^{1}$\thanks{Contact e-mail: d.clements@imperial.ac.uk},
G. Petitpas$^{2}$,
V. Asboth$^3$,
A. Conley$^{4}$\newauthor
I.~P{\'e}rez-Fournon$^{5,6}$
D. Riechers$^{7,8,9}$
\\
$^{1}$Imperial College London, Prince Consort Road, London SW7 2AZ, UK\\
$^{2}$Harvard-Smithsonian Center for Astrophysics, 60 Garden Street, Cambridge, MA 02138 \\ 
$^3$Department of Physics and Astronomy, University of British Columbia, 6224 Agricultural Road, Vancouver, BC V6T-1Z1, Canada\\
$^{4}$Center for Astrophysics and Space Astronomy 389-UCB, University of Colorado, Boulder, CO, 80309, USA\\
$^{5}$Instituto de Astrof\'{i}sica de Canarias, C/ V\'{i}a L\'{a}ctea, E-38200 La Laguna, Tenerife, Spain\\
$^{6}$Departimento de Astrof\'{i}sica, Universidad de La Laguna, E-38206, La Laguna, Tenerife, Spain\\
$^7$	Cornell University, 220 Space Sciences Building, Ithaca, NY 14853, USA\\
$^8$Max-Planck-Institut f\"ur Astronomie, K\"onigstuhl 17, D-69117 Heidelberg, Germany\\
$^9$Humboldt Fellow
}
\date{}

\pubyear{2019}

\begin{document}
\label{firstpage}
\pagerange{\pageref{firstpage}--\pageref{lastpage}}
\maketitle

\begin{abstract}
We present SMA observations at resolutions from 0.35 to 3 arcseconds of a sample of 34 candidate high redshift dusty star forming galaxies (DSFGs). These sources were selected from the HerMES {\em Herschel} survey catalogues to have SEDs rising from 250 to 350 to 500$\mu$m, a population termed 500-risers. We detect counterparts to 24 of these sources, with four having two counterparts. We conclude that the remaining ten sources that lack detected counterparts are likely to have three or more associated sources which blend together to produce the observed {\em Herschel} source. We examine the role of lensing, which is predicted to dominate the brightest (F500 $>$ 60mJy) half of our sample. We find that while lensing plays a role, at least 35\% of the bright sources are likely to be multiple sources rather than the result of lensing. At fainter fluxes we find a blending rate comparable to, or greater than, the predicted 40\%. We determine far-IR luminosities and star formation rates for the non-multiple sources in our sample and conclude that, in the absence of strong lensing, our 500-risers are very luminous systems with L$_{FIR} > 10^{13}$L$_{\odot}$ and star formation rates $> 1000$M$_{\odot}$/yr.

\end{abstract}





\section{Introduction}

The role and nature of dusty star forming galaxies (DSFGs) at high redshifts ($z>4$) is currently unclear. Observations with {\em Herschel} have identified an unexpectedly  large population of candidate high z DSFGs through colour selection in the three bands of its SPIRE instrument \citep{Dowell2014, Asboth2016, Ivison2016}. This selection is based on finding a rising spectral energy distribution (SED) from 250 to 500$\mu$m, leading to these sources being termed `500-risers'. Since the far-IR SED of a typical galaxy peaks at around 100$\mu$m in the rest frame, sources whose SED is still rising at 500$\mu$m are likely to lie at $z>4$.These 500-risers are found with densities on the sky of 3.3$\pm$0.8 deg$^{-2}$ down to 500$\mu$m fluxes of 30mJy \citep{Dowell2014}. Such sources, if at $z>4$, would have luminosities $>10^{13}$L$_{\odot}$ and star formation rates $>$1000$ M_{\odot}$/yr. They would thus be extreme objects and as such could represent the high luminosity end of a larger population of DSFGs at high redshift (see eg. Greenslade et al., 2019). However, spectroscopic redshifts have been obtained for only a tiny fraction of this population \citep[][]{Capak2008, Coppin2009, Daddi2009,Riechers2010, Cox2011, Capak2011, Combes2012, Walter2012, Riechers2013, Dowell2014, Yun2015, Ivison2016, Oteo2016, Asboth2016, Riechers2017, Zavala2017, Strandet2017, Fudamoto2017, Oteo2017b, Marrone2017} and so its properties are only weakly constrained, and there is disagreement on whether this population contributes significantly to the cosmic SFR-density (SFRD) above $z > 4$ \citep{Rowan-Robinson2016, Liu2017, Novak2017, yan2019} or not \citep{Michalowski2017, Koprowski2017}. It has even been suggested that current data are unable to tell the difference between models where high z DSFGs dominate the SFRD or not \citep{Casey2018}.

The nature of the individual 500-risers that make up this population is also in dispute. Simulations by \citet{Bethermin2017} suggest that most 500 $\mu m$ risers are in fact blends, where the brightest galaxy contributes on average 60\% of the total Herschel 500 $\mu m$ flux density, though this is also a function of 500 $\mu m$ flux, and above $S_{500} = 60~mJy$ they find almost all sources should be strongly lensed single galaxies (see Figure 6 of \citet{Bethermin2017}). Gravitationally lensed sources will appear as pairs of sources separated by a few arcseconds, or as arclike structures on similar scales (see \citet{Bussmann2013} for examples of lensed DSFGs observed by the SMA). \citet{Donevski2018} further suggest that confusion and instrumental noise also boost the number of 500-risers seen by {\em Herschel}. Such multiple sources could be chance alignments along the line of sight or might be physically associated systems within the same large scale structure. They might also be the result of galaxy interactions and mergers that could also trigger starburst activity leading to their high far-IR luminosity. Observations with single dish submm telescopes eg. \citet{Duivenvoorden2018} can go some way to addressing this since they can achieve somewhat higher angular resolutions than the 36 arcseconds of the {\em Herschel} 500$\mu$m beam. However, the idea that 500-risers are blends of multiple sources can best be tested with submm interferometry observations which can achieve angular resolutions of a few arcseconds or better, depending on configuration. Numerous inteferometric studies of {\em Herschel} detected sources have been performed \citep{Karim2013, Hodge2013, Swinbank2014, Simpson2015, Oteo2017a, Hill2018}, but most of these have targeted sources selected at 850 $\mu$m and not as 500-risers. 

The ideal test of the nature of the 500-riser population is to conduct submm interferomteric observations with resolutions of a few arcseconds or better. This will be able to determine the multiplicity fraction and, at higher resolutions, examine the role of gravitational lensing.
\citet{Oteo2017a} performed ALMA observations with a resolution of 0.12 arcseconds at 850$\mu$m on 500 $\mu m$ risers selected from \citep{Ivison2016}, reaching average sensitivities of 0.1mJy/beam, and with a primary beam FWHM of 17 arcseconds.
They find that 27 of their sample of 44 single dish sources (61\%) are single sources, whereas the remaining 16 sources (39\%) split into multiple galaxies.
Additionally, they find several sources with multiplicities $> 3$, and one source where a single Herschel source resolves into 5 galaxies in the ALMA maps. They also find that 18 of their targets are gravitationally lensed to some extent. These results are interpreted as reducing the tension between models and the 500-riser counts from {\em Herschel}, with flux boosting thanks to confusion and multiplicity accounting for the discrepancy. However, this is not a complete sample, so statistical conclusions are somewhat uncertain, and the small primary beam means that some of the 500$\mu$m flux measured by {\em Herschel} may be accounted for by further companion sources that lie outside the small ALMA beam.

It is thus clear that high resolution observations of 500 $\mu m$ risers can provide valuable constraints on models, particularly the multiplicity of such sources and the relative contribution of multiple sources to the single dish observed flux density.
There are strong predictions regarding the multiplicity of 500-risers, in particular that sources with S$_{500} > 60~mJy$, should almost entirely be lensed single galaxies, something not borne out by \citet{Oteo2017a}.
This paper describes Sub-Millimeter Array (SMA, \citet{Ho2004}) observations of 34 500-risers to determine the multiplicity of these sources. A future paper will also use this data to obtain cross identifications at other wavelengths (Clements et al. in prep). Objects are selected using a slightly less stringent method to those of  \citet{Oteo2017a} (see below). Similar to \citet{Oteo2017a}, our resulting set of observations does not represent a complete sample, but we expect it to be reasonably representative.
Results for a number of our sources have been presented elsewhere \citep{Riechers2013, Dowell2014}, but we here present the first uniform reduction of this data to examine the statistical properties of this population at resolutions of $\sim$ 2 arcsec.

The rest of this paper is structured as follows. In the next section we describe the sample of 500-risers observed. After that we describe the observations and data reduction, and present the results obtained. The subsequent section discusses the source properties derived from these observations and our other results. Finally, we draw our conclusions. Throughout this paper, we assume a concordance $\Lambda$-CDM cosmology, with H$_0 = 67.74$ km s$^{-1}$ $Mpc^{-1}$, $\Omega_{\Lambda} = 0.69$ and $\Omega_{m} = 0.31$.

\section{Sample Selection}

Our 500-riser targets are all selected from the HerMES survey \citep{Oliver2012} and include sources chosen from both preliminary and finally published maps and catalogs. All our sources were selected as candidate high redshift DSFGs based on their {\em Herschel} fluxes and colours, but their selection is heterogenous in the sense that there is no single consistent flux or colour criterion used. This means that while it is a representative sample of sources, rigorous statistical conclusions may be difficult to draw since it is not a complete sample. There are 34 sources in total, with 500 $\mu m$ flux densities ranging from 27 to 160 $mJy$, and an average of 67 $\pm$ 29 $mJy$. For consistency, we extract the SPIRE fluxes of our sources from the latest versions of the HerMES maps\footnote{Specifically data release 4} and catalogues. The methods used to detect sources and extract the final fluxes for these objects are described in \citet{Wang2014}.
The {\em Herschel} fluxes can be seen in Table \ref{table:targets}, which also includes the number of SMA detections (see Section 3). Further details of our sources, as well as details of the SMA observations, can be found in Table \ref{table:obs}. All sources were detected by {\em Herschel} at $>4\sigma$ with the majority at much higher levels of significance.
Colour selection was on the basis of flux rising to 500$\mu$m, ie. $F250 < F350 < F500$, the defining characteristic of a 500-riser,  based on the data available at the time of selection and observation. However, as can be seen, refinements to the data reduction in later versions of the maps and catalogues mean that a few sources, such as Lock2, HeLMS35, and HFLS5, can no longer strictly be classed as 500 $\mu m$ risers.
However, we include them since their initial selection was as a 500 $\mu m$ riser, and investigations by \citet{Donevski2018} and \citet{Bethermin2017} have shown that there can be a significant contamination of selected 500 $\mu m$ risers by flux boosted sources with $S_{500} \lessapprox S_{350}$.

\begin{table*}
\centering
\begin{tabular}{lrrrrrrrr} \hline
           Map &  Detections &   D$_{map}$Lim &   F250 &    F350 &    F500 &   E250 &   E350 &   E500 \\
           &   &  [$mJy$]  &   [$mJy$] &    [$mJy$] &    [$mJy$] &  [$mJy$] &   [$mJy$] &   [$mJy$] \\ \hline
      Bootes13 &         2 &                 4.8 &  54.1 &   80.6 &   81.8 &   7.8 &   8.1 &   8.3 \\
      Bootes15 &         0 &                 7.0 &  51.2 &   68.3 &   66.1 &   6.0 &   7.0 &   8.0 \\
      Bootes24 &         0 &                 9.0 &  28.7 &   49.7 &   55.8 &   6.0 &   7.0 &   8.0 \\
      Bootes27 &         0 &                 3.9 &  37.8 &    0.0 &   52.2 &   6.0 &   7.0 &   8.0 \\
      Bootes33 &         1 &                 4.1 &  50.0 &   60.9 &   57.0 &   8.1 &   8.7 &   8.0 \\
       Cosmos9 &         1 &                 2.8 &  20.6 &   32.9 &   47.6 &   6.1 &   5.9 &   6.9 \\
     HFLS1 &         1 &                10.4 &  51.6 &   82.4 &   82.9 &   9.8 &  10.3 &   9.7 \\
 HFLS3\_270 &         1 &                 7.1 &  15.2 &   33.4 &   52.8 &   6.7 &   8.0 &   9.3 \\
 HFLS3\_345 &         1 &                11.3 &  15.2 &   33.4 &   52.8 &   6.7 &   8.0 &   9.3 \\
     HFLS5 &         1 &                 6.2 &  24.8 &   53.0 &   45.8 &   8.4 &   9.2 &   9.1 \\
       HeLMS11 &         1 &                 7.7 &  83.9 &   89.6 &  110.4 &  15.6 &  14.9 &  17.5 \\
       HeLMS12 &         1 &                17.7 &  76.9 &  122.5 &  138.7 &  16.7 &  16.6 &  17.3 \\
       HeLMS28 &         1 &                16.8 &  91.2 &  137.8 &  160.5 &  14.9 &  14.2 &  16.9 \\
       HeLMS32 &         1 &                 8.1 &  36.0 &   61.5 &   72.4 &  16.0 &  14.8 &  16.9 \\
       HeLMS34 &         1 &                13.1 &  66.7 &   90.7 &  106.7 &  14.7 &  14.4 &  16.0 \\
       HeLMS35 &         2 &                11.1 &  54.2 &   91.0 &   68.6 &  15.9 &  15.5 &  17.9 \\
       HeLMS45 &         0 &                10.8 &  32.0 &   59.0 &   82.0 &  14.0 &  15.0 &  16.0 \\
      HeLMS493 &         0 &                 7.9 &  35.0 &   56.0 &   68.0 &  14.0 &  15.0 &  16.0 \\
       HeLMS62 &         1 &                10.6 &  36.0 &   30.6 &   41.0 &  27.4 &  17.0 &  17.1 \\
      HeLMS783 &         1 &                 6.2 &  42.0 &   46.6 &   61.0 &  15.0 &  15.0 &  17.0 \\
         LSW28 &         1 &                 7.0 &  41.0 &   65.1 &   66.9 &  11.8 &  11.9 &  13.2 \\
         LSW52 &         0 &                 4.3 &  16.3 &   33.0 &   40.2 &   6.0 &   7.0 &   8.0 \\
         LSW73 &         2 &                 3.9 &  19.3 &   27.4 &   35.7 &   8.4 &   8.6 &   7.7 \\
     Lock2\_270 &         2 &                 5.1 &  41.4 &   67.0 &   65.1 &   7.6 &   7.9 &   8.1 \\
     Lock2\_345 &         1 &                11.6 &  41.4 &   67.0 &   65.1 &   7.7 &   8.0 &   8.1 \\
     Lock5\_270 &         1 &                 4.6 &  17.5 &   38.1 &   44.1 &   8.3 &   9.1 &   8.9 \\
     Lock5\_345 &         1 &                11.0 &  17.5 &   38.1 &   44.1 &   8.3 &   9.1 &   8.9 \\
  Lock\_102 &         1 &                15.4 &  64.0 &  105.5 &  122.7 &   6.3 &   6.7 &   7.1 \\
        XMM-26 &         1 &                 7.3 &  45.7 &   67.8 &   81.6 &  11.9 &   9.6 &   8.8 \\
        XMM-65 &         0 &                 6.3 &   0.0 &   24.6 &   42.3 &   6.0 &   7.0 &   8.0 \\
        XMM-81 &         0 &                 6.8 &   0.0 &    9.4 &   28.5 &   6.0 &   7.0 &   8.0 \\
        XMM-M2 &         1 &                 9.4 &  44.5 &   54.1 &   67.6 &   8.7 &   9.9 &   9.4 \\
        XMM-M5 &         0 &                 9.1 &  26.6 &   44.0 &   46.1 &   6.0 &   7.0 &   8.0 \\
        XMM-M7 &         1 &                 9.8 &  39.7 &   60.3 &   57.6 &   7.4 &   8.2 &   9.3 \\
        XMM-R2 &         0 &                 9.2 &  23.0 &   32.8 &   40.3 &   6.0 &   7.0 &   8.0 \\
        XMM-R3 &         0 &                 9.0 &  22.3 &   34.9 &   37.6 &   8.0 &   8.7 &   9.7 \\
        XMM-R7 &         0 &                 8.1 &  14.3 &   24.9 &   37.3 &   9.8 &   9.3 &   9.0 \\ \hline
\end{tabular}
\caption[The 500-riser sample]{The {\em Herschel} fluxes of our targets, together with the number of detections in each of the maps, and the threshold used for detections on the dirty map, D$_{map}$Lim. SPIRE flux densities, in $mJy$, are given by extractions on the SPIRE maps at the positions of the sources. For sources where we have observations at multiple frequencies (HFLS3, Lock2 and Lock5), we split our maps into different frequencies and detect independently in both of them. HFLS3, Lock2 and Lock5, are observed at both 270 $GHz$ and 345 $GHz$, labelled as \_270 and \_345 respectively.}
\label{table:targets}
\end{table*}

\section{Observations and Data Analysis}

\subsection{Observations and Data Reduction}

Our targets were observed in a number of programmes at the SMA over the period from 2010 May 19 to 2015 May 7. The observations were made at a variety of different tunings but generally at frequencies close to either 345 or 265 GHz. The most common configuration used was Compact but observations of specific sources in SubCompact, Extended and, in two cases, Very Extended were also used. A mixture of full tracks, partial tracks and track sharing was used for these observations, resulting in different sampling of the $uv$ plane. A summary of the observations made, the number of antennae used, and the median opacity at 225GHz during them is given in Table \ref{table:obs}.

Each of the maps was reduced using the IDL based SMA data reduction package \textsc{MIR}\footnote{\url{https://github.com/qi-molecules/sma-mir}}, uniformly in the following way:
The data were manually inspected, and any place where individual integrations, spectral bands, baselines or antennae showed significant flux density offsets or phase changes were flagged and removed from the data.
Bandpass calibration used the nearest accessible bright flat spectrum QSO selected from the following sources: 3c279, 3c454.3, 3c84, and BL-Lac. Flux calibration sources were well characterised Solar System sources selected on the basis of availability from Titan, Uranus, Neptune, Callisto, and Ganymede. These are standard bandpass and flux calibrators used extensively for SMA observations.
Gain calibration was done periodically throughout an observation by observing one to two nearby reasonably bright QSOs with known flux densities at 270 and/or 345 $GHz$. These sources were chosen from the list of gain calibrators maintained by the SMA for this purpose. All calibration observations used the same tuning as the science observations.
The total calibration process results in flux uncertainties of $\sim$ 10\%.
After calibration, the continuum was generated by averaging all the spectral channels which were not flagged in the inspection step, and the maps were then exported to the data analysis package \textsc{MIRIAD} \citep{Sault2011} for further reduction.

We used the \textsc{MIRIAD} command \textsc{invert} to transform the visibilities into (dirty) maps, choosing a natural weighting (i.e. constant weight to all visibilities) since this maximises the signal to noise of any sources present.
We then \textsc{CLEAN}ed our maps, using 250 \textsc{CLEAN} iterations.
The noise level varies between maps, but on the \textsc{CLEAN}ed maps was typically around 1.5 $ \pm$ 0.5 $mJy$, with minimum and maximum values of 0.5 $mJy$ and 2.9 $mJy$ respectively. 
We extract sources and fluxes from the dirty maps rather than the \textsc{CLEAN}ed maps (see Section 4), and for the dirty maps our noise values had a mean of 2.2 $ \pm$ 0.9 $mJy$, with minimum and maximum values of 0.8 $mJy$ and 4.9 $mJy$ respectively.
The beamshape also varied between maps, and by assuming a Gaussian shape for the central part of the beam, had an average FWHM of 2.5 $\pm$ 0.8 arcsec.
At $z = 4$, the expected redshift of our 500 $\mu m$ risers, this corresponds to 17.5 kpc. The DSFGs observed by \citep{Oteo2017a} all have sizes less than 0.6 arcsec so it is unlikely that our sources will be resolved.
Multiple sources or signs of lensing within a $\sim 2.5$ arcseconds radius would also not necessarily be resolved. Experiments were performed using different weighting schemes and cleaning parameters, using both the data and simulations (see Section 3.2). The modelling discussed in Section 3.2 found that the process described above generally performed best at recovering the flux density of injected fake sources.
The noise levels and beamsizes in our maps are in reasonable agreement with what \citet{Hill2018} found in their recent SMA reduction of a number of SCUBA-2 selected sources. The resulting dirty maps, beamshapes, and CLEANed maps for all of our 500 $\mu m$ risers can be found in Appendix A. The CLEANED maps are not used for the scientific analysis but are included for presentational purposes and to show how the complex beam shapes in some of the dirty maps can mimic the presence of multiple sources.

\begin{table*}
    \centering
    \begin{tabular}{llllrllll} \hline
        Map     &           RA &           DEC &      Date &  Freq &         Config &    Project  & \# of & Median \\
                &             2000 &               2000&           &[$GHz$]&                &     Code    & Ants  & Opacity  \\ \hline
    HFLS1       &  17:08:17.67 &   58:28:45.11 &  05/19/10 &   270 &        Compact &  2010A-S035 &   7   &  0.05  \\
                &              &               &  03/14/11 &   265 &     SubCompact &  2010B-S042 &   7   &  0.11  \\
                &              &               &  03/23/11 &   265 &     SubCompact &  2010B-S042 &   6   &  0.11 \\
                &              &               &  08/02/11 &   265 &       Extended &  2011A-S053 &   8   &  0.09  \\
    HFLS3       &  17:06:47.69 &   58:46:23.88 &  05/12/10 &   270 &        Compact &  2010A-S035 &   7   &  0.08  \\
                &              &               &  03/14/11 &   265 &     SubCompact &  2010B-S042 &   7   &  0.11  \\
                &              &               &  03/23/11 &   265 &     SubCompact &  2010B-S042 &   6   &  0.11  \\
                &              &               &  08/02/11 &   265 &       Extended &  2011A-S053 &   8   &  0.09  \\
                &              &               &  08/03/11 &   345 &       Extended &  2011A-S053 &   8   &  0.09  \\
                &              &               &  07/06/11 &   336 &     SubCompact &  2011A-S053 &   7   &  0.08  \\
                &              &               &  09/06/11 &   336 &  Very Extended &  2011A-S053 &   8   &  0.06  \\
    Lock2       &  10:53:10.92 &   56:42:06.84 &  05/12/10 &   270 &        Compact &  2010A-S035 &   7   &  0.08  \\
                &              &               &  03/14/11 &   265 &     SubCompact &  2010B-S042 &   7   &  0.11  \\
                &              &               &  03/23/11 &   265 &     SubCompact &  2010B-S042 &   6   &  0.11  \\
    HFLS5       &  17:20:49.49 &   59:46:27.12 &  05/19/10 &   270 &        Compact &  2010A-S035 &   7   &  0.05  \\
                &              &               &  03/14/11 &   265 &     SubCompact &  2010B-S042 &   7   &  0.11  \\
                &              &               &  03/23/11 &   265 &     SubCompact &  2010B-S042 &   6   &  0.11  \\
                &              &               &  08/02/11 &   265 &       Extended &  2011A-S053 &   8   &  0.09  \\
    Lock5       &  10:51:32.09 &   56:36:17.64 &  05/19/10 &   270 &        Compact &  2010A-S035 &   7   &  0.05  \\
                &              &               &  03/14/11 &   265 &     SubCompact &  2010B-S042 &   7   &  0.11  \\
                &              &               &  03/23/11 &   265 &     SubCompact &  2010B-S042 &   6   &  0.11  \\
                &              &               &  02/14/12 &   345 &       Extended &  2011B-S020 &   8   &  0.07  \\
    XMM\_M2     &  02:25:15.10 &  -02:47:09.20 &  12/08/11 &   345 &        Compact &  2011B-S038 &   8   &  0.07  \\
    XMM\_R7     &  02:29:29.32 &  -04:42:17.00 &  12/08/11 &   345 &        Compact &  2011B-S038 &   8   &  0.07  \\
    XMM\_M7     &  02:26:44.76 &  -03:25:05.50 &  12/08/11 &   345 &        Compact &  2011B-S038 &   8   &  0.07  \\
    XMM\_M5     &  02:18:56.74 &  -04:35:44.90 &  12/09/11 &   345 &        Compact &  2011B-S038 &   8   &  0.08  \\
    XMM\_R3     &  02:26:05.19 &  -03:18:28.10 &  12/09/11 &   345 &        Compact &  2011B-S038 &   8   &  0.08  \\
    XMM\_R2     &  02:17:43.86 &  -03:09:11.20 &  12/09/11 &   345 &        Compact &  2011B-S038 &   8   &  0.08  \\
    XMM\_26     &  02:25:45.31 &  -02:59:16.10 &  11/27/12 &   265 &        Compact &  2012B-S018 &   7   &  0.12  \\
    XMM\_81     &  02:21:57.82 &  -04:12:25.60 &  11/27/12 &   265 &        Compact &  2012B-S018 &   7   &  0.12  \\
    XMM\_65     &  02:18:57.91 &  -03:35:05.60 &  11/27/12 &   265 &        Compact &  2012B-S018 &   7   &  0.12  \\
    LSW\_28     &  11:01:27.29 &   56:19:31.40 &  11/27/12 &   265 &        Compact &  2012B-S018 &   7   &  0.06  \\
    LSW\_73     &  10:56:19.61 &   56:53:50.60 &  11/27/12 &   265 &        Compact &  2012B-S018 &   7   &  0.06  \\
    LWS\_52     &  10:55:48.89 &   57:33:57.24 &  11/27/12 &   265 &        Compact &  2012B-S018 &   7   &  0.06  \\
    Bootes\_27  &  14:38:45.14 &   33:22:31.80 &  03/19/13 &   265 &        Compact &  2012B-S018 &   8   &  0.06  \\
    Bootes\_33  &  14:31:32.11 &   34:21:16.60 &  03/19/13 &   265 &        Compact &  2012B-S018 &   8   &  0.06  \\
    Bootes\_13  &  14:35:43.51 &   34:47:42.70 &  03/19/13 &   265 &        Compact &  2012B-S018 &   8   &  0.06  \\
   Lock\_102$^*$&  10:40:50.64 &   56:06:53.84 &  03/02/13 &   302 &        Compact &  2012B-S078 &   8   &  0.04  \\
                &              &               &  01/19/13 &   302 &       Extended &  2012B-S078 &   7   &  0.03  \\
                &              &               &  02/03/13 &   302 &  Very Extended &  2012B-S078 &   8   &  0.07  \\
    HeLMS\_28   &  00:44:09.94 &   01:18:28.40 &  07/02/13 &   265 &        Compact &  2013A-S005 &   6   &  0.13  \\
    HeLMS\_12   &  00:52:58.52 &   06:13:18.80 &  07/02/13 &   265 &        Compact &  2013A-S005 &   6   &  0.13  \\
    HeLMS\_34   &  00:22:21.16 &  -01:55:21.70 &  07/02/13 &   265 &        Compact &  2013A-S005 &   6   &  0.13  \\
    HeLMS\_493  &  23:54:11.79 &  -08:29:12.00 &  07/15/13 &   265 &        Compact &  2013A-S005 &   6   &  0.10  \\
    HeLMS\_45   &  00:03:04.39 &   02:40:49.80 &  07/15/13 &   265 &        Compact &  2013A-S005 &   6   &  0.10  \\
    HeLMS\_62   &  00:55:17.18 &   02:04:01.20 &  07/15/13 &   265 &        Compact &  2013A-S005 &   6   &  0.10  \\
    HeLMS\_32   &  23:24:35.09 &  -05:24:51.80 &  07/09/13 &   265 &        Compact &  2013A-S005 &   6   &  0.09  \\
    HeLMS\_783  &  00:41:29.86 &  -02:47:48.10 &  07/09/13 &   265 &        Compact &  2013A-S005 &   6   &  0.09  \\
    HeLMS\_11   &  00:29:36.26 &   02:07:09.80 &  07/09/13 &   265 &        Compact &  2013A-S005 &   6   &  0.09  \\
    Bootes\_15  &  14:40:09.66 &   34:37:55.70 &  06/30/14 &   346 &        Compact &  2014A-S092 &   8   &  0.06  \\
    Bootes\_24  &  14:36:21.30 &   33:02:29.00 &  06/30/14 &   346 &        Compact &  2014A-S092 &   8   &  0.06  \\
    Cosmos\_9   &  10:02:49.11 &   02:32:57.40 &  11/28/14 &   231 &        Compact &  2014A-S092 &   7   &  0.08  \\
    HeLMS\_35   &  00:21:15.55 &   01:32:58.60 &  05/07/15 &   345 &        Compact &  2014A-S092 &   7   &  0.06  \\ \hline
    \end{tabular}
    \caption[Target properties]{The positions, dates, frequencies, array configurations, and project codes for the observations of HerMES selected 500 $\mu m$ risers, ordered roughly by the date of observation. A list of the SPIRE fluxes at the positions of these sources in given in Table \ref{table:targets}.
$^*$ Also described in \citep{Wardlow2013} with the source name HLock09.}
    \label{table:obs}
    \end{table*}

\subsection{Source Extraction}

We extract sources from the dirty maps on the basis of SNR with the noise calculated from the whole dirty map, without correction for the primary beam response. We also examined using noise calculated within 1 - 2 beam FWHM radii of the source instead of the global value.

To determine the reliability of source extraction, we perform the following steps.
First, for each of our maps, we create 1,000 fake maps filled with Gaussian noise.
These are then convolved with the beam for that map, and this fake dirty map is scaled until the noise properties match those of the original dirty map (i.e. same mean and standard deviation).
This provides 1,000 realisations of each map but without sources. The same extraction method as on the real data is then applied to these empty maps.

The best extraction method will minimise the number of fake sources detected, whilst simultaneously minimising the SNR threshold used for detection, so as to maximise the depth of our our catalog.
We found that using the global SNR measured on the dirty map performed best, with only 0.09\% fake sources detected at $> 5\sigma$, compared to the local measurement on the dirty map, which had 2\% fake sources detected at $>5\sigma$. We also experimented with source extraction on the \textsc{CLEAN}ed maps but found this to have even higher rates of fake sources.

The average number of fake sources detected as a function of SNR varies significantly between maps. This is likely because the synthesised beams can be very different. We thus select a SNR cutoff for each map such that there is an average of 0.1 fake sources per map. This SNR cutoff varies from map to map, ranging from 3.54 to 3.86 $\sigma$,  but is on average 3.75$\sigma$, where $\sigma$ is the global noise on the dirty map. Over all our 34 sources, we would thus expect 3.4 $\pm$ 1.8 spurious, low SNR sources.

We test the completeness of our extraction method by injecting fake sources into the real maps and then attempting to recover them.
To do this, we copy each of our observed maps 1,000 times, and inject a point source convolved with the beam with a flux distributed uniformly between 0 and 30 $mJy$.
We require the injected sources to be at least 5 arcseconds away from any detected source in the map.
These maps then have sources extracted and we determine what fraction of injected sources are recovered within one synthesised beamwidth of the injected position.
We find significant variations between the maps, with maps that have more complex beams (e.g. HeLMS28) having worse source recovery than maps with better UV coverage (e.g. Bootes13). We find on average that we recover 65.3\% of sources.
Of the remaining 34.7\%, most are faint, with only 6.6\% having flux densities $> 10~mJy$. These are usually in maps where a bright source is already present, raising the global noise.

We also use our injected sources to examine two alternative methods for extracting flux densities:  using Gaussian fitting, or fitting the beam to the dirty map at the position of the peak pixel, in addition to measuring the peak flux from the map.
We find that extracting the peak flux from the map minimised the difference between injected and extracted flux densities, compared to both Gaussian and beam fitting.
We note that we only injected point sources, so the alternate methods may perform better for resolved sources.
We find good agreement between injected and recovered fluxes for flux densities $> 10 - 15~mJy$.
Below this, the injected flux is within a few standard deviations of the noise of most of our maps, and while the recovered flux is generally within 2$\sigma$ of the injected value, the (recovery - injected) to injected flux density ratio diverges.

Where we have additional extended or very extended observations (EXT or VEXT) of a given source, we find that maps combining these with compact or subcompact (COM or SUBCOM) observations produced significantly lower recovered fluxes compared to compact (or subcompact) observations on their own.
The reason for this remains unclear, but phase decoherence or partially resolving the source in extended configurations are likely explanations. We thus extract flux densities only using compact or subcompact observations. In Figure \ref{fig:allmaps}, these are indicated with the addition of ``\_COM\_'' to the map name.

\section{Results}

\subsection{Detected Sources}

We apply the source detection and extraction methods described above to our maps of 500-risers, finding 26 sources associated with our 34 {\em Herschel} selected 500-risers. Three of these sources are detected at two different frequencies in the SMA maps. The number of SMA sources associated with each 500-riser is given in Table \ref{table:targets} while the individual SMA sources are detailed in Table \ref{table:detections}.

\begin{table*}
\centering
\begin{tabular}{lrrrrrrl} \hline
       Name &         RA &       DEC &  Frequency &   Flux &  Flux\_Err &  SNR$^*$ &     $z$ \\
       &          2000&        2000&  [$GHz$] &   [$mJy$] &  [$mJy$] &   &     \\ \hline
      Bootes13ID\_1 &  218.93331 &  34.79416 &    265 &   10.4 &      1.25 &              8.32 &       \\
      Bootes13ID\_2 &  218.93281 &  34.79787 &    265 &   9.35 &      1.25 &              7.48 &       \\
      Bootes33ID\_1 &  217.88329 &  34.35276 &    265 &   6.23 &      1.08 &              5.77 &       \\
       Cosmos9ID\_1 &  150.70492 &   2.54869 &    231 &   3.62 &      0.66 &              5.48 &       \\
     HFLS1\_ID\_1 &  257.07377 &  58.47777 &    265 &  15.8 &      2.87 &              5.50 &  4.29$$ \\
 HFLS3\_270\_ID\_1 &  256.69914 &  58.77330 &    270 &  18.4 &      1.36 &             13.5 &  6.34 \\
  &  256.69926 &  58.77330 &    336 &  24.6 &      1.79 &              13.7 &  6.34 \\
     HFLS5\_ID\_1 &  260.20754 &  59.77442 &    270 &  13.3 &      1.32 &              10.1 &  4.44 \\
       HeLMS11ID\_1 &    7.40011 &   2.12036 &    265 &   10.0 &      2.07 &              4.83 &       \\
       HeLMS12ID\_1 &   13.24405 &   6.22189 &    265 &  38.6 &      3.80 &              10.1 &  4.37 \\
       HeLMS28ID\_1 &   11.04294 &   1.30615 &    265 &  34.4 &      3.75 &              9.17 &  4.17 \\
       HeLMS32ID\_1 &  351.14402 &  -5.41560 &    265 &  12.6 &      1.90 &              6.63 &       \\
       HeLMS34ID\_1 &    5.58661 &  -1.92247 &    265 &  26.0 &      3.25 &              8.00 &  5.16 \\
       HeLMS35ID\_1 &    5.31452 &   1.54934 &    345 &  19.1 &      2.61 &              7.32 &       \\
       HeLMS35ID\_3 &    5.31506 &   1.55261 &    345 &   13.0 &      2.61 &              4.98 &       \\
       HeLMS62ID\_2 &   13.82280 &   2.06797 &    265 &   11.9 &      2.91 &              4.09 &       \\
      HeLMS783ID\_1 &   10.37588 &  -2.79840 &    265 &   7.86 &      1.72 &              4.57 &       \\
         LSW28ID\_1 &  165.36617 &  56.32630 &    265 &  19.2 &      1.71 &             11.2 &       \\
         LSW73ID\_1 &  164.08463 &  56.90126 &    265 &   4.19 &      1.00 &              4.19 &       \\
         LSW73ID\_2 &  164.07629 &  56.89420 &    265 &   4.38 &      1.00 &              4.38 &       \\
     Lock2\_270ID\_1 &  163.29714 &  56.70437 &    270 &   9.07 &      0.94 &              9.65 &       \\
     Lock2\_270ID\_2 &  163.29509 &  56.70212 &    270 &   5.68 &      0.94 &              6.04 &       \\
     &  163.29503 &  56.70220 &    345 &  13.1 &      3.0 &              4.36 &       \\
     Lock5\_270ID\_1 &  162.88248 &  56.60535 &    270 &   8.0 &      0.80 &              10.0 &  3.36 \\
     &  162.88288 &  56.60542 &    346 &  14.2 &      2.92 &              4.86 &  3.36 \\
  Lock\_102\_ID\_1 &  160.21099 &  56.11496 &    302 &  72.1$^*$ &      2.99 &             24.1 &  5.29 \\
        XMM-26ID\_1 &   36.43590 &  -2.98359 &    265 &   7.62 &      1.92 &              3.96 &       \\
        XMM-M2ID\_1 &   36.31398 &  -2.78553 &    345 &  14.4 &      2.51 &              5.73 &       \\
        XMM-M7ID\_1 &   36.68632 &  -3.41962 &    345 &   13.3 &      2.55 &              5.21 &       \\ \hline
\end{tabular}
\caption[Properties of the detected sources]{Properties of our 29 detected sources. The redshift, if known, is given in the final column. In three cases sources are detected at two different frequencies. $^*$ SNR extracted from the dirty map. Redshift sources are as follows: HFLS1, 3, 5 and Lock\_102 from \cite{Riechers2013}; Lock5 from \cite{Dowell2014}; HeLMS34 from \cite{Asboth2016}, HeLMS12 and 28 from Riechers et al. (in prep), and \cite{Duivenvoorden2018}. $^*$ Lock\_102\_ID\_1 is our most extended source so this flux is derived not from the point source extraction applied to all of the other sources but instead extracts flux from a 4 arcsecond diameter aperture on the COM configuration map. The flux extracted from the EXT map in a similar way is 68.8mJy, which is within $\sim 1\sigma$, but we use the COM flux to minimise the possibility of flux being resolved out. The point source extracted flux for this source is 55.0mJy.}
\label{table:detections}
\end{table*}

Of our 34 500-risers, 12 (35\%) have no detections with the SMA, 18 (53\%) have a single detection, and 4 (12\%) have two detections, indicating that at these flux density limits and resolutions, most 500 $\mu m$ risers appear to be single sources.
The sample size is fairly small, but there is no evidence at these flux densities to suggest any correlation between the number of detections and the detection threshold used in the map, with the two, one, and no detection subsets having mean detection thresholds of $6.2 \pm 3.3$, $9.4 \pm 4.0$, and $7.6 \pm 2.0~mJy$ respectively.
There are also similar numbers of detections for the 345 $GHz$ and 270 $GHz$ maps, though there are only 13 maps at 345 $GHz$ and 24 at 270 $GHz$.
The flux densities of detected sources range between 3.62 to 35.7 $mJy$ at 270 $GHz$, with a mean of 12.3 $mJy$, and between 13.0 and 72.1 $mJy$ at 345 $GHz$, with a mean of 22.96 $mJy$.

Of the 17 SPIRE sources with S$_{500} > 60~mJy$, which \citet{Bethermin2017} predict to be almost entirely single sources, we find 12 that appear to be single sources, 3 have multiple detections, and 2 have zero detections. 
One of the multiple sources, Lock2, has one counterpart at 345 $GHz$ but two at 270 $GHz$. 
Curiously, the source detected at both 345 $GHz$ and 270 $GHz$ is the fainter of the two sources at 270 $GHz$, with only 5.68 $\pm$ 0.94 $mJy$ at 270 $GHz$ compared to the other source which has 9.07 $\pm$ 0.94 $mJy$ but no counterpart at 345 $GHz$.
As we discuss later, the maps where we detect no sources are likely be made up of multiple faint sources below the detection limit.
This means that of our 17 SPIRE sources with S$_{500} > 60~mJy$, 5 (30\%) are multiple galaxies.
Of the 18 sources in \citet{Oteo2017a} with S$_{500} > 60~mJy$, they find 6 (33\%) that are resolved into multiple galaxies, in good agreement with our SMA result. Such a high fraction would appear to disagree with \citet{Bethermin2017}, who suggest such sources are almost entirely lensed single sources.

We can only say with certainty that the maps where we detect 2 sources are definitely multiples; in maps where we detect a single source, there may be a secondary fainter source, and in maps where we detect no sources, we can at best put upper limits to any detection.
For maps with one detection, the ratio between the brightest (detected) source and any secondary source can be a maximum of $\frac{S_{Lim}}{S_{Detected}}$, where $S_{Lim}$ is the upper limit for detection, and $S_{Detected}$ is the flux density of the detected source.
We find that in this case, any secondary source would have a flux density \textit{at most} between 98\% and 28\% that of the detected source, with an average of (63 $\pm$ 19)\%.
These upper limits are reasonably consistent with \citet{Donevski2018}, who find that the brightest source that makes up a Herschel detected 500 $\mu m$ riser contributes around 60\% of the 500 $\mu m$ flux density.
However, at the current sensitivities our estimates remain upper limits, though we note that in the four maps where we have two detections the brighter source contributes on average 60\% to the total flux density from both sources.
For now, the simplest solution to the maps where we detect only one source is that this sole source is responsible for the SPIRE flux, and we move ahead under this assumption.
The same cannot be said for those maps where we detect no sources.

\subsection{Maps Without Detections}

In twelve maps, we detect no sources at all.
This is surprising: all but one of these maps have Herschel 500 $\mu m$ flux densities $> 40~mJy$, while Bootes15, HeLMS493 and HeLMS45 have 500 $\mu m$ flux densities $> 60~mJy$. The detection threshold in these dirty maps is $\sim 7-10~mJy$, making it difficult to explain these non-detections as single sources that are below our detection threshold.
Examining Figure \ref{fig:allmaps}, there is some evidence of a \textsc{CLEAN}ed source in HeLMS493, with a CLEANed signal to noise ratio of 5.61 and a flux density of 9.85 $\pm 1.76~mJy$, but this source does not pass our threshold in the dirty map, as it has a signal to noise ratio of only 2.1 and a flux density of 4.65 $\pm 2.18~mJy$.
Numerous ($>2$) DSFGs which appear as single sources in the Herschel maps are known to be multiple at higher resolutions \citep{Oteo2017b, Oteo2018}.
To determine how many sources would reproduce the typical SPIRE flux in the maps where we do not have any SMA detections, we take the ALESS average SED \citep{daCunha2015} at redshifts between 4 and 6, and normalise its flux density at 500 $\mu m$ to 60 $mJy$.
If a single source were responsible for the SPIRE flux, we find that it should have 345 $GHz$ and 270 $GHz$ flux densities of $\sim$ 24-38 $mJy$ and 14-25 $mJy$ respectively, and that we should have detected the source responsible.
If the flux were equally split between two sources, we would expect them to have SMA flux densities of $\sim$ 12-19 and 7-12 $mJy$ respectively; low, but, given the detection limits in Bootes15 and HeLMS493, still detectable.
This is not a detailed analysis, but it suggests that in at least some of the maps where we do not detect any sources there may be $\geq 3$ DSFGs separated by $\sim$ 2 - 20 arcsec (the SMA and SPIRE beam sizes respectively). This raises the intriguing possibility that such sources may be protocluster cores, with numerous far-IR luminous sources spread over a $\sim$20 arcsec region, such as the Distant Red Core \citep{Oteo2018} or SPT2349-56 \citep{Miller2018}.
We have thus begun a programme to obtain deeper observations of these `blank field' sources both at high resolution with the SMA, and with the JCMT to determine integrated fluxes (Cairns et al., in prep).

\section{Discussion}

\subsection{The Physical Properties of 500-risers}

To determine the physical parameters of our sources, we take the measured SPIRE and SMA flux densities, and, following literature convention \citep{Blain2002, Magnelli2012, Bianchi2013, Casey2014}, fit them to a single temperature modified blackbody function of the form
\begin{equation}  
    \label{chap3:eq:MBB}
    S_{\nu} \propto (1 - exp(-\tau_{\nu})) B_\nu(T),
\end{equation}
where $S_{\nu}$ is the observed flux density at frequency $\nu$, $\tau_{\nu}$ = $(\frac{\nu}{\nu_{0}})^{\beta}$ gives the optical depth at frequency $\nu$, $\nu_{0}$ is the frequency at which the optical depth equals unity, and $B_{\nu}(T) = B_{\nu}(\nu, T_{dust})$ is the Planck function at temperature $T$.
$\beta$, the spectral index, is usually assumed to be $\beta = 1.5$ - 2 for SMGs \citep{Blain2002, Casey2014}. Since it is unclear how to distribute SPIRE flux between multiple SMA sources, we do not include these sources in this analysis. 

We use the affine invariant Markov Chain Monte Carlo \citep{Goodman2010} ensamble sampler \textsc{Python} package  \textsc{emcee} \citep{ForemanMackey2013} to fit this model to our data, with the following uninformative priors for our parameters: 0 $ < z \leq $ 12, $T_{CMB}(z) \leq T_{dust} \leq 80$, $1 \leq \beta \leq 3$, 1 $\mu m$ $ \leq c/\nu_{0} \leq 1$ $mm$, and $-2 \leq log_{10}(a) \leq 2$, where $T_{CMB}$ gives the CMB temperature at redshift $z$, and $c$ gives the speed of light.
For numerical stability, at each sample we first normalise to the 500 $\mu m$ observation, and allow the normalisation $a$ to vary from there.
In this model, there are therefore 5 parameters to fit: the redshift $z$, the average dust temperature $T_{dust}$, the spectral index $\beta$, $\nu_{0}$, and an overall normalisation parameter $a$. Where a redshift is known the value of $z$ in the fit is fixed to this value.
We run our sampler with 100 walkers and for 3000 steps and, after manually checking to ensure the model has burnt in, throw away the initial 300 steps as a burn in phase. The dust temperature and redshift that emerge from such fits are strongly degenerate, so these values do not provide any direct physical insight into our DSFGs. However, the far-IR luminosity, dust mass and star formation rate that can be derived from these fits suffer much less from this degeneracy (see e.g. \cite{Greenslade2019}). The far-IR luminosity (defined as the integrated luminosity from 42.5 - 122.5 $\mu m$) is calculated by integrating the fitted spectral energy distributions (SEDs) between these two rest frame frequencies. This is done for all samples on each source to produce the posterior luminosity distribution of each source marginalised over all redshifts. The dust mass is then calculated following \cite{Riechers2013} using
\begin{equation}
\label{chap3:eq:dmass}
    M_{dust} = S_{\nu}D_{L}^2[(1+z) \kappa_{\nu}B_{\nu}(T)]^{-1}\tau_{\nu}[1-exp(-\tau_{\nu})]^{-1},
\end{equation}
where S$_{\nu}$ gives the rest-frame flux density at 125 $\mu m$, D$_{L}$ is the luminosity distance, $\kappa_{\nu}$ is the mass absorption coefficient and is assumed to be $\kappa_{\nu} = 2.64~m^2~kg^{-1}$ at 125 $\mu m$ \citep{Dunne2003}, and other symbols are defined as before.
Finally, the FIR luminosity is converted to a SFR following \citet{Riechers2013}, using a Chabrier stellar IMF \citep{Chabrier2003}.
These results can be found in Table \ref{table:derived} where we show the median, 14th and 86th percentiles of these distributions for each of our sources. The resulting SED fits compared to the observational data are shown in Figure \ref{fig:seds}.

\begin{table*}
\centering
\begin{tabular}{lllll}\hline
   Name &                   $log_{10}(L_{FIR}/L_{\odot})$ &                  $log_{10}(M_{Dust}/M_{\odot})$ &                    $\beta$ &                     SFR [$M_{\odot} yr^{-1}$]\\ \hline
  Bootes33ID\_1 &  13.4$^{+0.26}_{-0.86}$ &  8.75$^{+0.76}_{-0.52}$ &  2.29$^{+0.43}_{-0.47}$ &  2800$^{+1400}_{-1900}$ \\
   Cosmos9ID\_1 &  13.3$^{+0.26}_{-0.85}$ &  8.64$^{+0.98}_{-0.55}$ &  2.55$^{+0.31}_{-0.45}$ &  2200$^{+1200}_{-1400}$ \\
 HFLS1 &  13.5$^{+0.07}_{-0.13}$ &  9.28$^{+0.24}_{-0.13}$ &  1.80$^{+0.56}_{-0.46}$ &    3500$^{+700}_{-600}$ \\
         HFLS3 &  13.5$^{+0.2}_{-0.21}$ &  9.05$^{+0.28}_{-0.13}$ &  1.88$^{+0.69}_{-0.60}$ &    3500$^{+400}_{-400}$ \\
 HFLS5 &  13.3$^{+0.16}_{-0.14}$ &  9.14$^{+0.29}_{-0.20}$ &  1.56$^{+0.67}_{-0.39}$ &    2200$^{+700}_{-500}$ \\
   HeLMS11ID\_1 &  13.5$^{+0.33}_{-0.76}$ &  8.96$^{+0.80}_{-0.52}$ &  2.32$^{+0.45}_{-0.56}$ &  3500$^{+2200}_{-2400}$ \\
   HeLMS12ID\_1 &  13.7$^{+0.12}_{-0.13}$ &  9.62$^{+0.23}_{-0.14}$ &  1.49$^{+0.68}_{-0.35}$ &  5500$^{+1400}_{-1000}$ \\
   HeLMS28ID\_1 &  13.7$^{+0.09}_{-0.09}$ &  9.64$^{+0.18}_{-0.17}$ &  1.66$^{+0.60}_{-0.40}$ &   5500$^{+1200}_{-900}$ \\
   HeLMS32ID\_1 &  13.5$^{+0.38}_{-0.82}$ &  8.93$^{+0.82}_{-0.47}$ &  2.06$^{+0.59}_{-0.63}$ &  3500$^{+3000}_{-2300}$ \\
   HeLMS34ID\_1 &  13.7$^{+0.17}_{-0.07}$ &  9.26$^{+0.17}_{-0.14}$ &  1.51$^{+0.66}_{-0.36}$ &  5500$^{+1800}_{-1100}$ \\
   HeLMS62ID\_2 &  13.6$^{+0.65}_{-0.93}$ &  9.04$^{+0.87}_{-0.56}$ &  1.86$^{+0.72}_{-0.61}$ &  4400$^{+7400}_{-2200}$ \\
  HeLMS783ID\_1 &  13.3$^{+0.41}_{-0.88}$ &  8.79$^{+0.75}_{-0.48}$ &  2.02$^{+0.65}_{-0.65}$ &  2200$^{+2600}_{-1500}$ \\
     LSW28ID\_1 &  13.6$^{+0.37}_{-1.2}$ &  9.20$^{+0.70}_{-0.44}$ &  1.50$^{+0.64}_{-0.36}$ &  4400$^{+5000}_{-3100}$ \\
         Lock2 &  13.4$^{+0.18}_{-0.84}$ &  8.70$^{+1.07}_{-0.56}$ &  2.57$^{+0.3}_{-0.48}$ &   2800$^{+800}_{-2700}$ \\
         Lock5 &  13.8$^{+0.15}_{-0.15}$ &  9.39$^{+0.39}_{-0.36}$ &  2.06$^{+0.64}_{-0.68}$ &  6900$^{+1100}_{-1800}$ \\
      Lock\_102 &  13.8$^{+0.05}_{-0.02}$ &  10.21$^{+0.73}_{-0.42}$ &  2.01$^{+0.69}_{-0.76}$ &  6900$^{+900}_{-400}$ \\
    XMM-26ID\_2 &  13.6$^{+0.21}_{-0.74}$ &  8.82$^{+0.99}_{-0.55}$ &  2.65$^{+0.25}_{-0.40}$ &  3900$^{+2200}_{-3000}$ \\
    XMM-M2ID\_1 &  13.4$^{+0.22}_{-0.84}$ &  8.75$^{+0.81}_{-0.49}$ &  2.51$^{+0.35}_{-0.56}$ &  4400$^{+1200}_{-1600}$ \\
    XMM-M7ID\_1 &  13.4$^{+0.21}_{-0.70}$ &  8.66$^{+0.90}_{-0.48}$ &  2.41$^{+0.40}_{-0.58}$ &  4400$^{+1000}_{-1500}$ \\ \hline
\end{tabular}
\caption[Derived properties of the non-multiple detected sources]{The derived properties of our sources from fitting single modified blackbodies, {\em uncorrected for lensing $\mu$}. Lock2 was fitted only using the source detected at both 270 and 345 $GHz$.}
\label{table:derived}
\end{table*}

\begin{figure*}
    \centering
    \includegraphics[width=15cm]{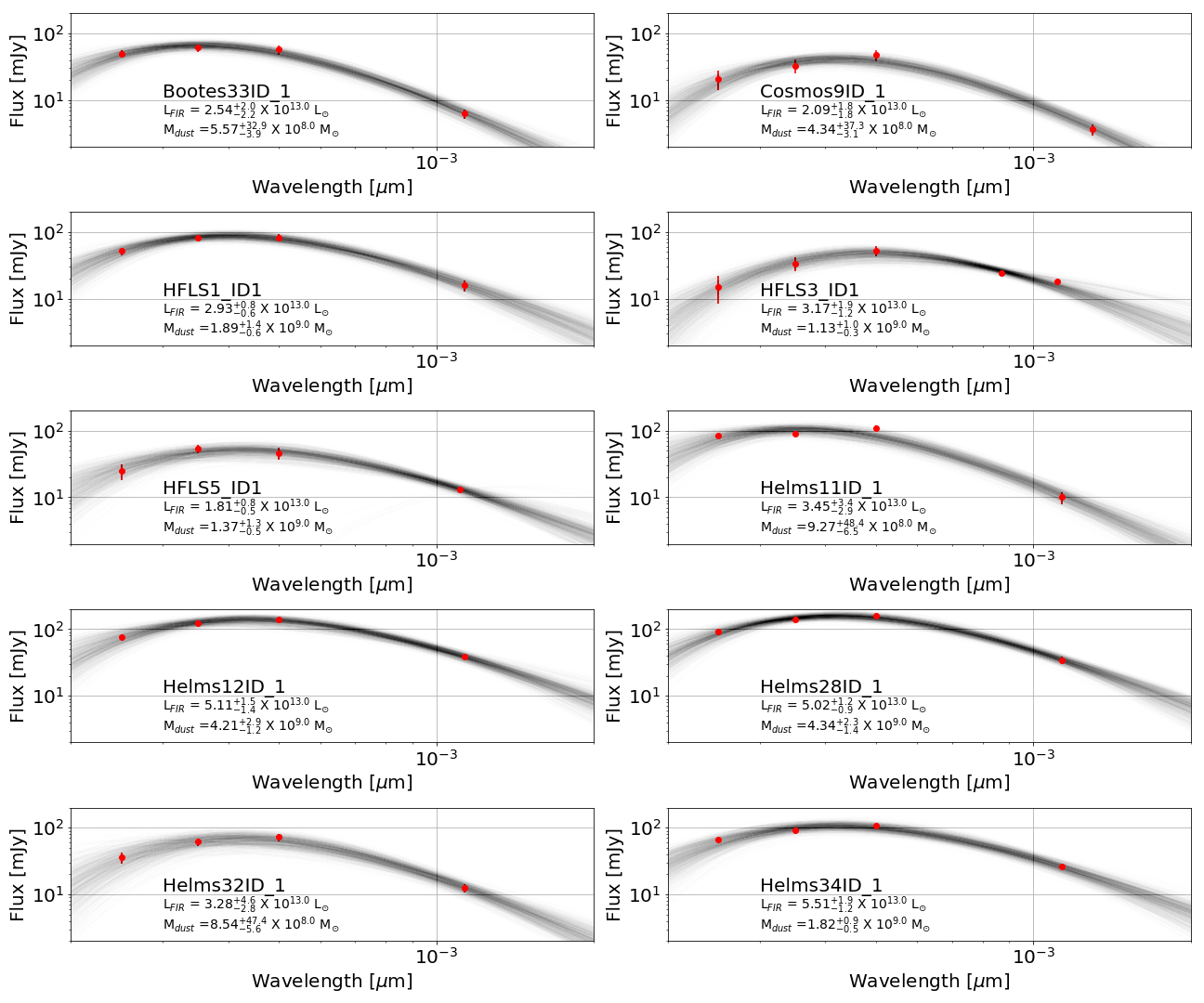}
    \caption[Fitted models to the SPIRE and SMA flux densities]{Sampled SED fits to the SPIRE and SMA flux densities measured for sources where we only detect a single counterpart to the SPIRE source. The black lines show 1,000 samples after fitting, while the red points give the flux densities and error bars to the observed photometry of each source. Names, estimated FIR luminosities and estimated dust masses are provided as text on each of the plots. The poor fit for Lock\_102 results from the presence of the [CII] 157$\mu$m line in the bandpass of these observations, contaminating and enhancing the submm flux derived from our SMA data (see Perez-Fournon et al., in prep).}
    \label{fig:seds}
\end{figure*}
\begin{figure*}
    \centering
    \includegraphics[width=15cm]{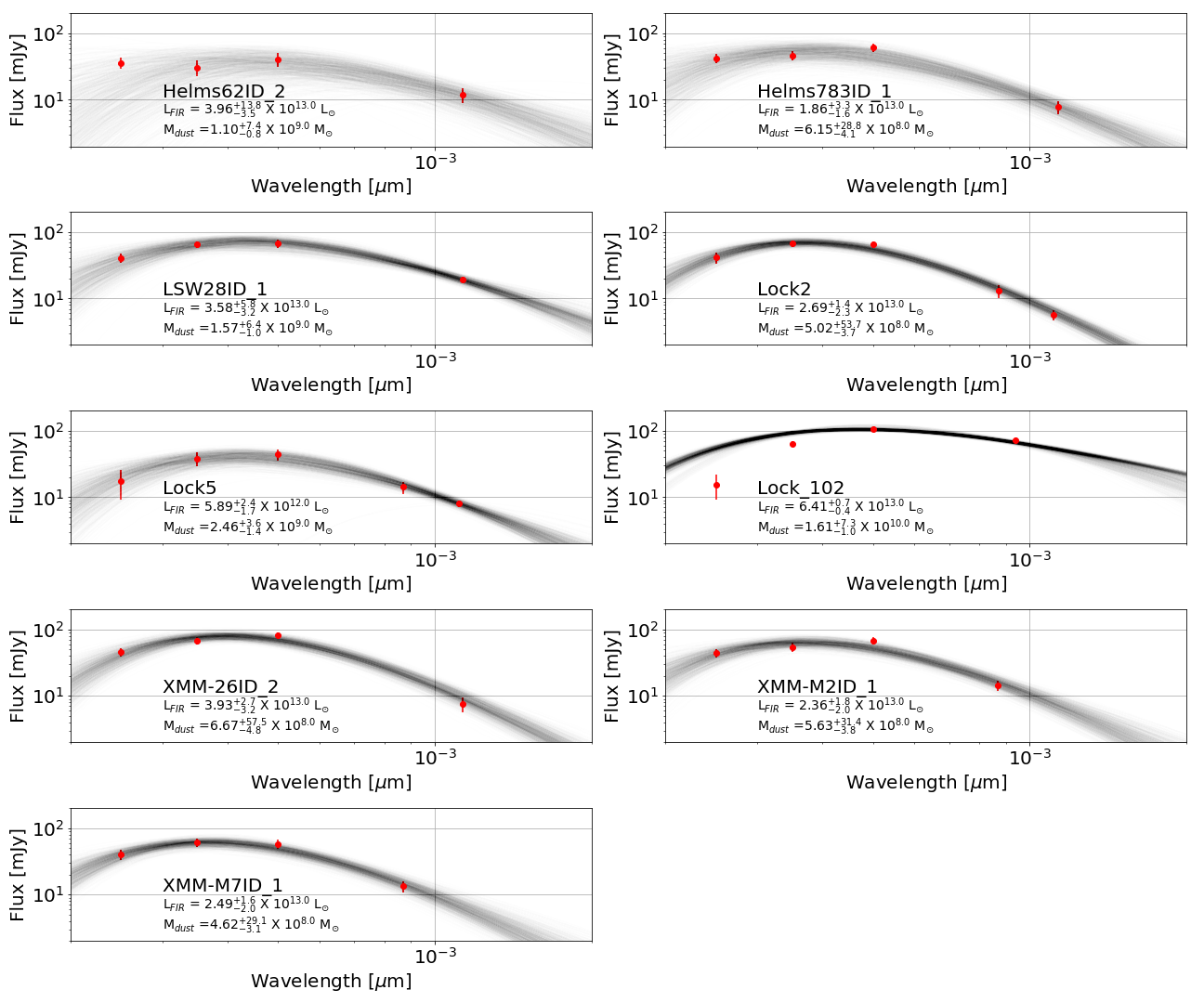}
   \contcaption{}
\end{figure*}

As can be seen from Table \ref{table:derived}, those sources where we detect a single SMA counterpart all have luminosities $> 10^{13}$M$_{\odot}$, leading to their classification as HLIRGs. While lensing may have boosted the derived luminosity of some sources, a modest lensing amplification at the level of 2 - 3, as is the case for HFLS3 \citep{Cooray2014} would still leave most of our sources as HLIRGs. In the absence of any AGN contribution to their high far-IR luminosity,  the derived SFR for these objects is in excess of 1000 $M_{\odot}$/yr, with some sources potentially having SFRs in excess of 5000 $M_{\odot}$/yr, leading to their classification as Extreme Starbursts \citep{RRobinson2018}, a class of object that cannot be accounted for by any current model of galaxy formation. While the error bars on these derived properties are large, with many lying within 1$\sigma$ of HLIRG luminosity, a number, such as HeLMS34, are well above this value. The simulations used by \citet{Bethermin2017} to determine the effects of source clustering on single dish instruments limited the maximal starburst to 1000 $M_{\odot}$yr$^{-1}$. Similar to the higher resolution ALMA observations of \citet{Oteo2017a}, we here show that sources beyond this limit exist in the 500-riser population, so it is likely that the results of the clustering simulations are in need of revision at the brightest fluxes. Further observations at complementary wavelengths, including the optical, X-ray and far-IR spectroscopy, will be needed to determine what role AGN play in powering these sources.

\subsection{The Role of Lensing}

\begin{figure*}
    \centering
    \includegraphics[width=\textwidth]{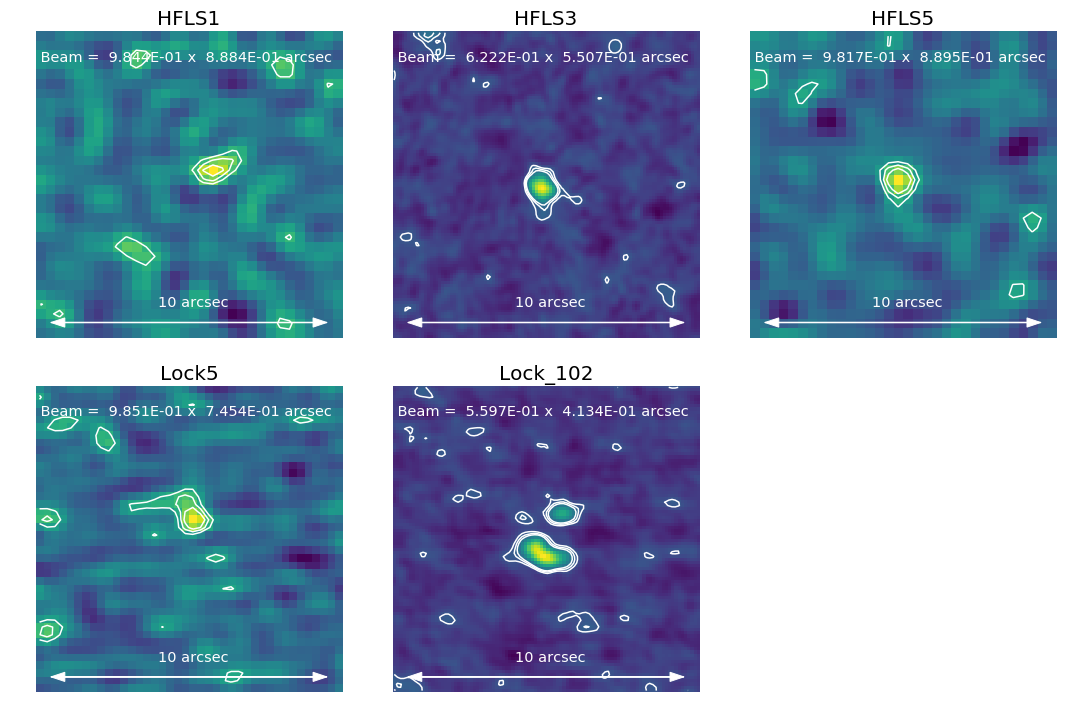}
    \caption[Extended SMA observations]{Extended (all maps) and very extended (HFLS3 and Lock\_102) only) observations on maps where these configurations are available. All maps are 10 arcsec $\times$ 10 arcsec, and orientated so that north is up and east is to the left. Contours show 2, 3, 4 and 5$\sigma$ levels. The beam is approximated as a Gaussian, with semi-major and semi-minor axis given in the text at the bottom of each image.}
    \label{fig:Lenses}
\end{figure*}

One of our sources is known to be strongly lensed (Lock\_102), and another (HFLS3) is weakly lensed with a magnification of 1.3. Seventeen of our sources have S$_{500} > 60~mJy$, which \citet{Bethermin2017} suggests should all be lensed galaxies.
However, we only have extended array observations for 5 of our sources (HFLS1, HFLS3, HFLS5, Lock5 and Lock\_102), and the 2 - 4 arcsecond beamsize in the compact and subcompact configurations lack the resolution necessary to provide a definitive test of lensing.
In Figure \ref{fig:Lenses}, we show those sources where we do have extended or very extended configuration observations. The synthesised beams for these range from 1 to 0.35 arcseconds across.
Only one of these sources, Lock\_102, clearly shows the arc-like structure indicating strong lensing, which is confirmed by observations at other telescopes. These observations were at a resolution of $\sim$ 0.35 arcseconds.
The other known lens, HFLS3 with a magnification $\mu$ = 2.2 $\pm$ 0.3 \citep{Cooray2014}, appears reasonably compact, with some sign of extension to the north, but no obvious indications of lensing. These observations were also at a resolution of $\sim 0.35$ arcseconds.
Similar suggestions of extension are seen in HFLS1 and Lock5 (resolutions of 1 and 0.8 arcseconds respectively) whilst HFLS5 (resolution 1 arcsecond) appears compact, indicating a diameter $< 6.9$ kpc.

\citet{Oteo2017a} in their ALMA examination of 500 $\mu m$ risers find 40\% of their sources display some lensing signature (arcs, elongation or rings). 
This is a significantly higher fraction than we find here, but the superior resolution and sensitivity offered by ALMA, with an average beam FWHM of only 0.12 arcsec, makes this kind of analysis much easier. Nevertheless, both the current work and \citet{Oteo2017a} find that the lensing fraction for 500-risers with flux densities $\geq$ 60mJy seems to be much lower than predicted by \citep{Bethermin2017}.

\subsection{The Role of Blended Sources}

Several authors have claimed, mostly through simulations, that most Herschel sources are multiple galaxies blended together in the large Herschel beam \citep{Cowley2015, Scudder2016, Bethermin2017, Scudder2018}.
\citet{Cowley2015} focused on 850 $\mu m$ selected sources and \citet{Scudder2016, Scudder2018} focused on 250 $\mu m$ selected sources, so their results are not well matched to the current results on 500-risers. In contrast, \citet{Bethermin2017} did look at 500$\mu$m selected sources. They suggest that on average 60\% of a S$_{500} < 60~mJy$ Herschel flux density at 500 $\mu m$ comes from a single source, while sources brighter than 60mJy should largely be strongly lensed single sources.
Our observations show a more complicated picture.
Of the 17 sources in our sample with F500$>$60mJy, three are found to have two counterparts and three are found to have no counterparts at all, suggesting that they are blends of three or more sources. This means that 35\% of the sources in our sample with flux densities $>$60 mJy at 500$\mu$m do not match the expectation that they should be single, strongly lensed sources. Instead they are blends of multiple components.
At brighter fluxes, S$_{500} > 100~mJy$, they do appear to be single sources, with one, Lock\_102, being a confirmed lens. At somewhat fainter fluxes, where lensing is still expected to dominate, we have counter-examples such as Bootes13, which resolves into 2 sources (S$_{500} = 82~mJy$) and Bootes15, where we detect no counterparts despite a flux density of  S$_{500} =  66~mJy$. At fainter fluxes, S$_{500} < 60~mJy$, where 40\% of sources are expected to be multiple, we find that of 17 sources, seven are single sources, one resolves into two components, and nine have no detection, suggesting that they are made up of multiple components, suggesting a multiplicity fraction of 60\%.

\begin{figure}
    \centering
    \includegraphics[width=8cm]{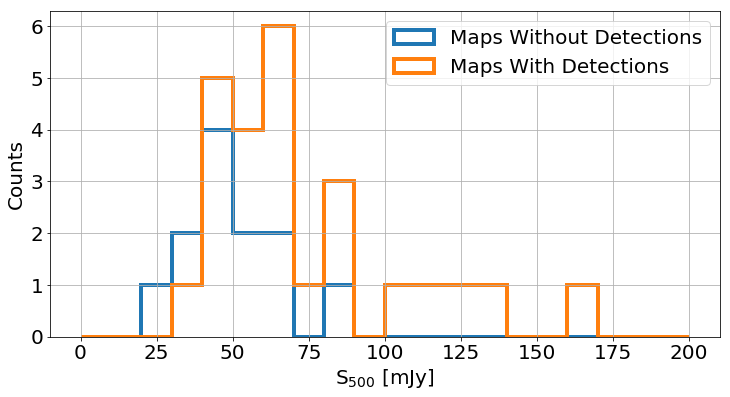}
    \caption[Detections as a function of Herschel flux]{The Herschel 500 $\mu m$ flux density of a source, as a function of whether a source is detected in the SMA map (orange histogram) or not (blue histogram). Bins are of width 20 $mJy$}
    \label{fig:500det}
\end{figure}

In Figure \ref{fig:500det}, we plot a histogram of the 500 $\mu m$ flux density of our sources split by whether we do or do not detect counterparts in our SMA maps.
The maps where we do not detect any sources are generally fainter at 500 $\mu m$. This tends to support the conclusion of \citet{Bethermin2017} that sources below 60mJy are more likely to be blends. However, with a likely multiplicity fraction of 60\% our data suggests that the blended fraction is likely to be higher than the predicted 40\%.

Comparing our results to those of \citet{Oteo2017a} is complicated by the fact that our observations were taken with a range of sensitivities, frequencies and configurations. Our best sensitivities would be able to detect at least 2 separate sources in all of their multiple sources, while our best resolution would be able to resolve all the multiple components in all of their targets. More typical observations, however, with a resolution of 2 arcsec and a 345 GHz equivalent sensitivity of 2mJy, would miss many of the multiple components found in \citet{Oteo2017a}, with most being classified as single sources and three being classified as blank fields. This suggests that the true multiplicity fraction may be even higher than we report above, implying a greater disagreement with the models. Differences in target selection for the two samples, however, mean that this is not yet a solid conclusion.  Our comparison with \citet{Oteo2017a} also suggests that deeper observations of our blank field sources will likely uncover further detections, something that our continuing SMA observations have already demonstrated for some sources (Cairns et al., in prep).

\section{Conclusions}

We have conducted SMA interferometric observations of 34 candidate high redshift ($z>4$) DSFGs selected from {\em Herschel} survey data. The sources were selected to be 500-risers, ie. sources whose redshifted far-IR dust SED is still rising in the SPIRE 500$\mu$m band. This has been found to be a good indicator of high redshift. Theoretical models \citep{Bethermin2017} have ascribed the unexpectedly high number counts of such 500-riser sources to the effects of gravitational lensing for bright sources (flux densities $>$60 mJy) and blending of multiple sources in the large {\em Herschel} beams at lower flux densities. Our observations, at resolutions from 0.35 to 3 arcseconds, detect counterparts to 22 of our 34 sources, with those sources lacking counterparts likely to be made up of three or more blended sources. We examine the rates of detection of counterparts above and below the 60mJy 500$\mu$m flux density break above which \citep{Bethermin2017} suggests all sources should be strongly lensed, and below which they suggest 40\% should be blends. We find that bright ($>$ 60 mJy) 500$\mu$m sources are a more diverse population with $\sim$35\% being blends rather than single sources. Fainter ($<$ 60 mJy) sources are found to be blends rather more often then predicted. We use the {\em Herschel} fluxes and our new SMA fluxes to determine the SED properties of the single sources we have detected, and then to measure their far-IR luminosity and SFR. We find that all our targets are HLIRGs, ie. have L$_{FIR} > 10^{13}$L$_{\odot}$, and with SFRs in excess of $>$ 1000M$_{\odot}$/yr if the effects of gravitational lensing are ignored. The 500-riser population thus seems likely to represent galaxies undergoing major bursts of star formation at very high redshifts. Such high luminosity objects were not considered in the \citet{Bethermin2017} models. Determining the broader role of this population in galaxy formation, and whether, as is likely for any reasonable luminosity function, there is an underlying much larger population of lower luminosity high redshift DSFGs, remain important tasks for far-IR and submm astronomy.

\section{Acknowledgements}

The authors wish to recognise and acknowledge the very significant cultural role and reverence that the summit of Mauna Kea has always had within the indigenous Hawaiian community.  We are most fortunate to have the opportunity to conduct observations from this mountain.
The Submillimeter Array is a joint project between the Smithsonian Astrophysical Observatory and the Academia Sinica Institute of Astronomy and Astrophysics and is funded by the Smithsonian Institution and the Academia Sinica.
This research has made use of data from the HerMES project (http://hermes.sussex.ac.uk/). HerMES is a Herschel Key Programme utilising Guaranteed Time from the SPIRE instrument team, ESAC scientists and a mission scientist.
The HerMES data was accessed through the Herschel Database in Marseille (HeDaM - http://hedam.lam.fr) operated by CeSAM and hosted by the Laboratoire d'Astrophysique de Marseille.
This research made use of Astropy, a community-developed core Python package for Astronomy \citep{Astropy2013}.
and of the Starlink Table/VOTable Processing Software, TOPCAT \citep{Taylor2005}.
This research also made use of NASA's Astrophysics Data System Bibliographic Services.
DLC and JG acknowledge support from STFC, in part through grant numbers ST/N000838/1 and ST/K001051/1. DR acknowledges support from the National Science Foundation under grant number AST-1614213 and from the Alexander von Humboldt Foundation and the Federal Ministry for Education and Research through a Humboldt Research Fellowship for Experienced Researchers. IPF acknowledges support from the Spanish Ministerio de Ciencia, Innovacion
y Universidades (MICINN) under grant numbers ESP2015-65597-C4-4-R and ESP2017-86852-C4-2-R. Many thanks to Joe Cairns and Tai-An Cheng for useful comments.




\bibliographystyle{mnras}
\bibliography{library} 

%



\begin{appendix}

\section{SMA Images}

We here present the dirty maps, beam shapes and cleaned maps for all at targets at all their observed frequencies. All images are 20 arcseconds on a side. North is up and east is to the left.

\begin{figure*}
    \centering
    \includegraphics[width=0.8\textwidth]{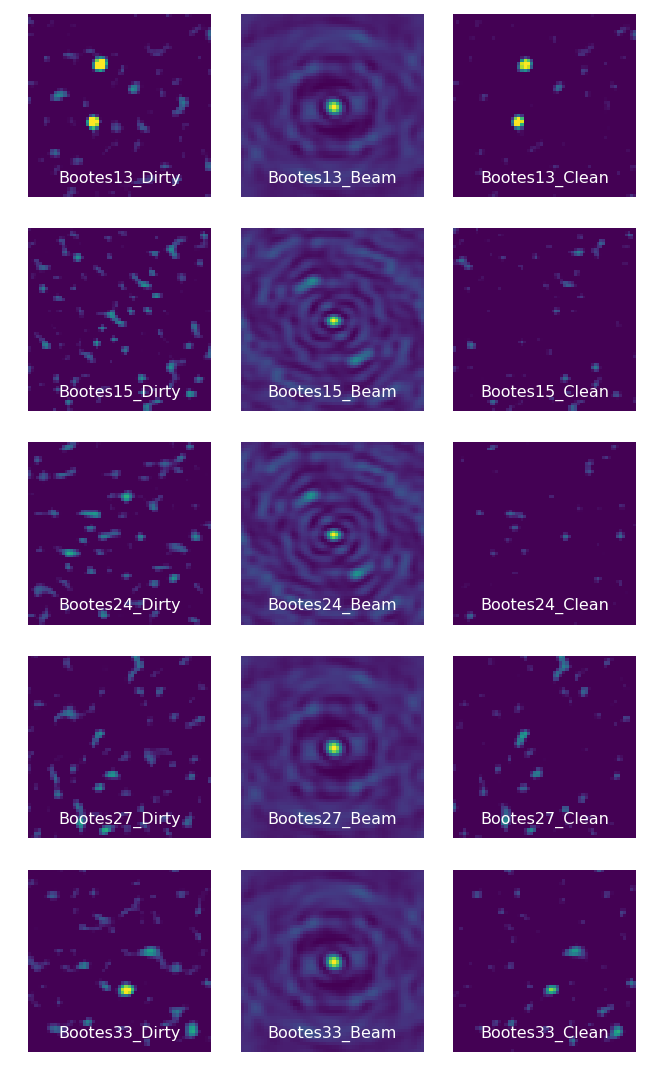}
    \caption[SMA maps and beam]{Images of the dirty (left), beam (middle) and \textsc{CLEAN}ed (right) maps of our 34 sources. The colour scheme in the dirty and clean maps ranges ranges between 1 and 5$\sigma$, using the rms on the dirty map. Maps which include the \textsc{\_COM\_} label indicate where we have only included the visibilities from the compact or subcompact array configurations (see text for more details).}
    \label{fig:allmaps}
\end{figure*}
\begin{figure*}
    \centering
    \includegraphics[width=0.8\textwidth]{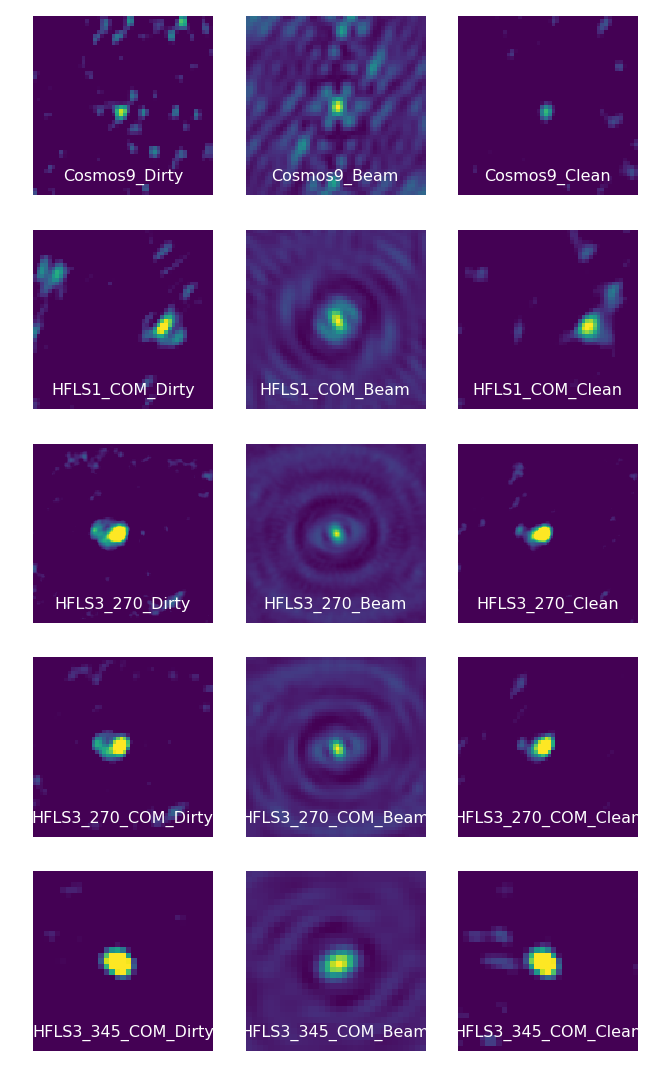}
   \contcaption{}
\end{figure*}
\begin{figure*}
    \centering
    \includegraphics[width=0.8\textwidth]{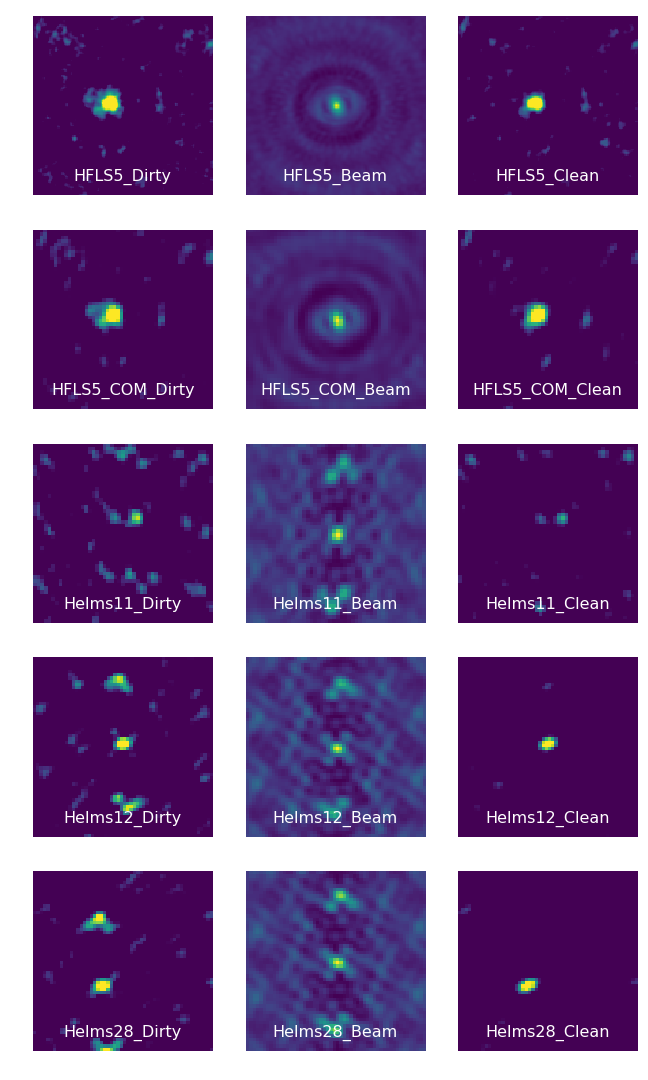}
   \contcaption{}
\end{figure*}
\begin{figure*}
    \centering
    \includegraphics[width=0.8\textwidth]{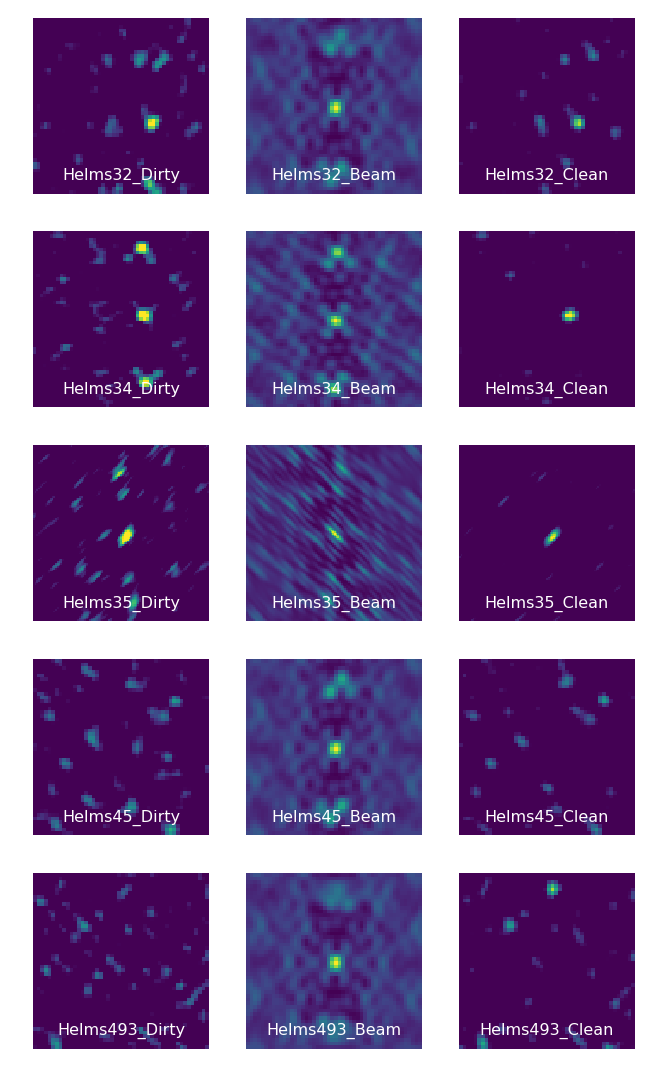}
   \contcaption{}
\end{figure*}
\begin{figure*}
    \centering
    \includegraphics[width=0.8\textwidth]{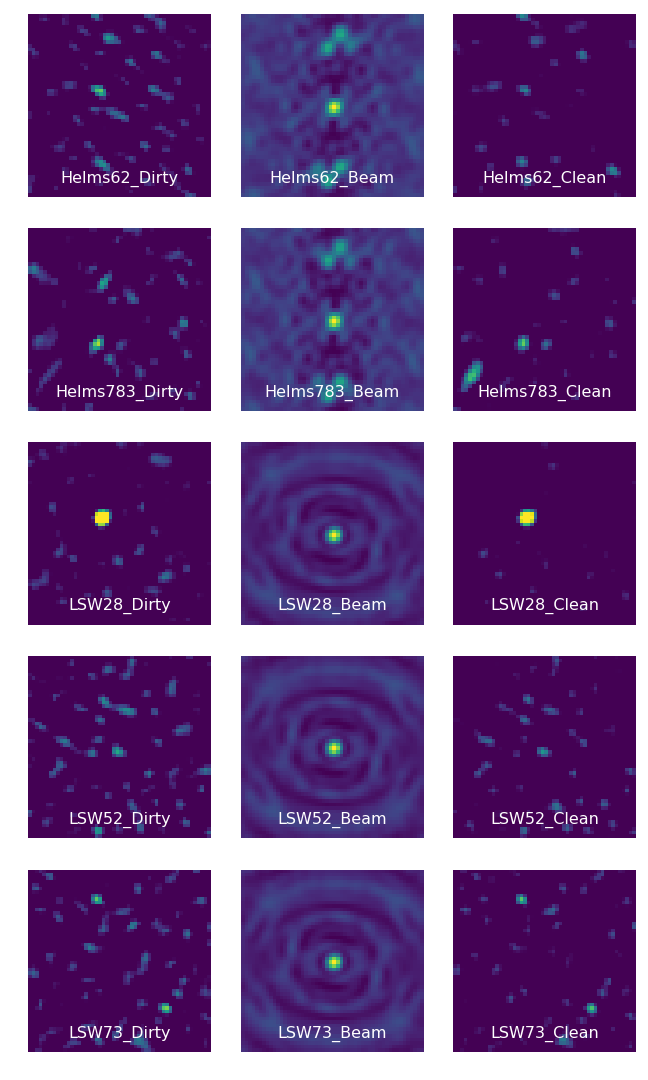}
   \contcaption{}
\end{figure*}
\begin{figure*}
    \centering
    \includegraphics[width=0.8\textwidth]{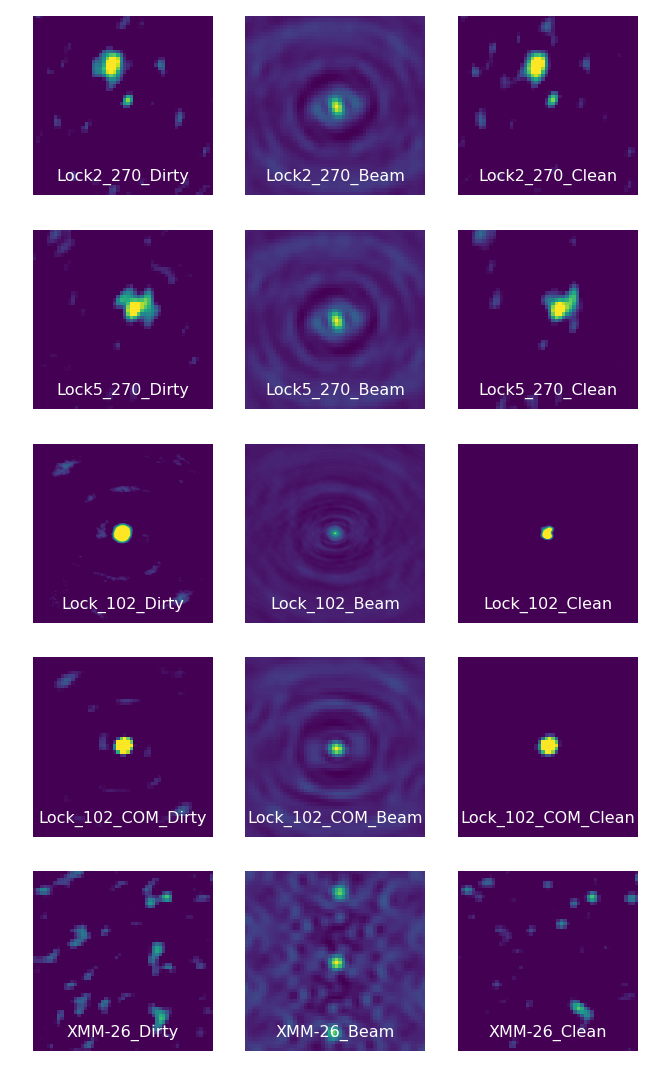}
   \contcaption{}
\end{figure*}
\begin{figure*}
    \centering
    \includegraphics[width=0.8\textwidth]{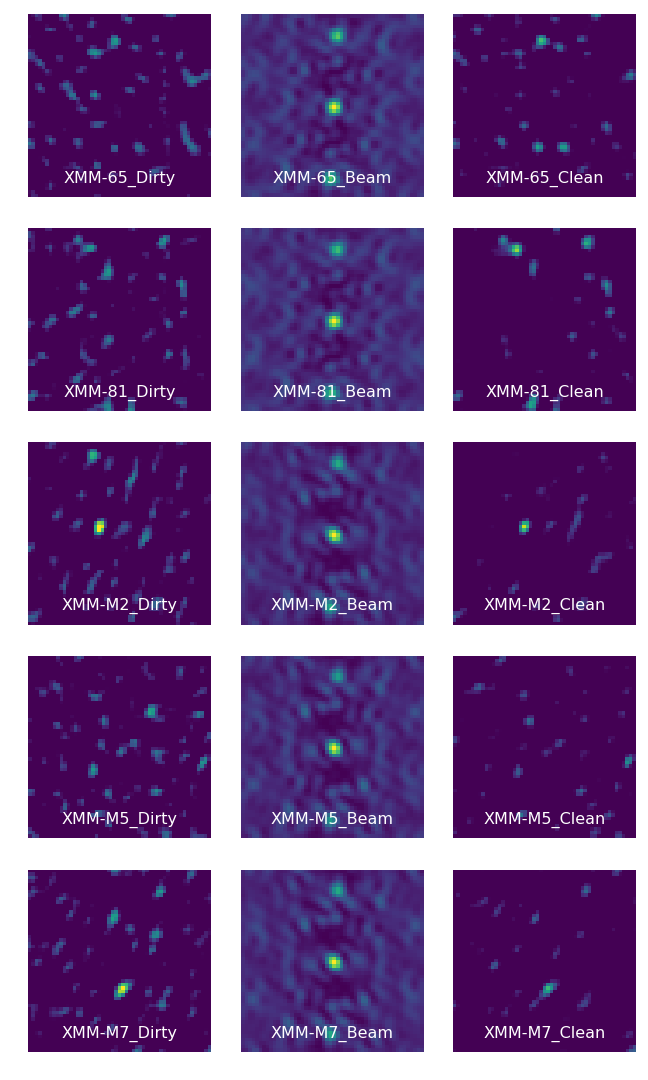}
   \contcaption{}
\end{figure*}
\begin{figure*}
    \centering
    \includegraphics[width=0.8\textwidth]{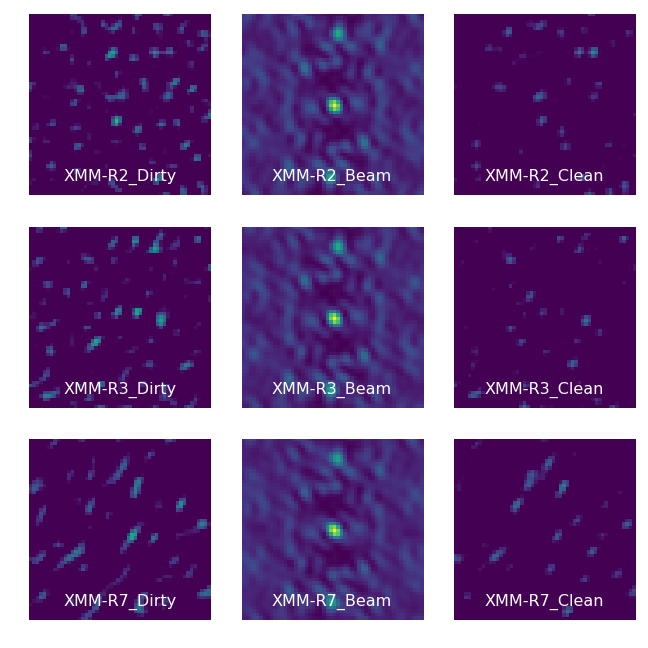}
   \contcaption{}
\end{figure*}

\end{appendix}

\bsp	
\label{lastpage}
\end{document}